# Local Articulation Points in Complex Networks

Senbin Yu[1], Liang Gao[1, 2, *], Rongqiu Song[1]

1. Institute of Transportation Systems Science and Engineering, MOE Key Laboratory of Urban Transportation System Theory and Technology, State Key Laboratory of Rail Traffic Control and Safety, and Center of Cooperative Innovation for Beijing Metropolitan Transportation, Beijing Jiaotong University, Beijing 100044, China
2. Yinchuan Municipal Bureau of Big Data Management and Service, Yinchuan 750011, P. R. China

**Abstract:** An articulation point (AP) is any node whose removal increases the number of connected components of a graph. There is no doubt that this kind of node which occupies a non-ignorable fraction of real-world networks plays a key role in ensuring the connectivity. However, we should not thus neglect the impacts of non-APs nodes. In this paper, we define a local AP (LAP) whose removal will increase the number of connected components within its $r$-step neighborhood. Through investigating the fraction of LAPs in forty-five real networks, we find a critical proportion $s_{cr}$, which is equal to 0.5 ($s = r/D$, $D$ is the diameter of a network), and this result can also be turned out in ER networks. In addition, we present a unique advantage of LAPs in dismantling networks under the process of targeted attack, compared with APs, which provide another way of thinking to improve the calculation efficiency of APs and design better targeted attack strategy of network destruction.

## 1 Introduction

A node is an articulation point (AP) or a cut-point if its removal disconnects the network or increases the number of connected components of the network [1, 2]. APs can be easily identified using a linear-time algorithm based on depth-first search [3]. Many real networks, such as infrastructure networks, protein interaction networks, and terrorist communication networks, have a non-ignorable fraction of APs [4].

APs are vital to many real networks. For example, air traffic networks and power grids networks would be exposed to serious risks if they were disrupted or attacked [6, 7]. In wireless sensor networks, failures of APs will disconnect the network and block data transmission process among each component [8-10]. In protein-protein interaction networks, lethal mutations enrich in the groups of highly connected proteins, where are also APs [11]. Identification of APs helps us better solving prior challenging works, such as calculating determinants of large matrices [12], assessing network stability [14], routing in delay tolerant networks [15], identifying structural hole spanners [16], and addressing vertex cover problem on large graphs, which is also known as the NP-complete problem [17].

Recently, a general framework about APs has been developed in the field of complex network science, and it has solved some key issues pertinent to APs [4, 5].

However, despite the importance of APs in ensuring the global robustness and connectivity of real-world networks, we could not neglect the significance of remaining nodes which take up a larger fraction of a network. For example, food webs networks [18] nearly have no APs, thus we should consider how to identify the importance of nodes in this kind of network. Although there is no difference among APs, the removal of different-positioned APs could have varied impacts on the network [4, 19]. Thus, we believe that each node has its own importance for the connectivity in a network, and the differences between them can be found by various observation scales.

Based on this hypothesis, in this paper, we define a local AP (LAP) whose removal will increase the number of connected components within its $r$-step neighborhood. Then, by analyzing $n_{LAP}$ in real networks, we find a critical proportion $s_{cr}$ at 0.5, which is also valid in ER networks. Further, we find that LAPs perform more efficient on the collapse process of giant connected component (GCC) in some real networks under targeted attack, especially in infrastructure networks and social networks, compared with the AP-targeted attack (APTA) strategy.

The structure of this paper is organized as follows. In section 2, the concept of LAP and a new network dismantling strategy based on LAP removal are introduced. In section 3, $n_{LAP}$ and $s_{cr}$ of real-networks and ER networks are calculated, and the experiment results of network destruction process based on LAP removal strategy are presented. In the last part, the experiment results and the conclusion are summarized.

## 2. Definition and method

### 2.1 LAP

A simple approach to distinguish a node is an AP or not is to remove it and see whether its removal disconnects the network. As shown in Fig. 1, if we remove the green node in the network (a), then the remaining nodes disconnect with each other, thus the green node is an AP. Comparatively, in the sample network (b), which contains network (a), the green node that located in the center is not an AP. However, we cannot deny its importance of connectivity in the network (a) and (b). In addition, every yellow node in the network (b) is also an AP in their own 1-step neighborhood subgraph, but they are not defined as APs. Therefore, each node may be an AP where has its importance for ensuring the connectivity in their own local environment. In other words, the classic concept of AP cannot cover some influential nodes in maintaining the local connectivity.

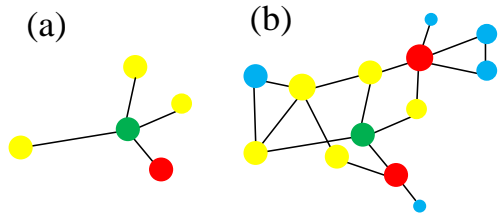

Fig. 1. Schematic networks. (a) There are 5 nodes in the network, and the green node is an AP. (b) The network contains 13 nodes and 17 edges, and the APs are in red.

Therefore, we define a new type of AP called local AP (LAP). It is a node whose removal would increase the number of connected components within its $r$-step neighborhood. In brief, it is a measure that considers whether a node is important from the local perspective. According to the definition of LAP, the green node and five yellow nodes in Fig. 1(b) are LAPs within their 1-step neighborhood, then we can name them as $LAP^1$. Obviously, the two APs in red can also be regarded as $LAP^1$s, $LAP^2$s, $LAP^3$s or $LAP^4$s. In general, the limit of $r$ is no more than the diameter ($D$) for a network, in case there exists more than one biconnected component in a network, where $D$ is measured by the maximum biconnected component. In a word, an AP belongs to one kind of $LAP^r$ no matter what value of $r$ is and vice versa. The set of $LAP^{r1}$s includes the $LAP^{r2}$s, if $r2$ is larger than $r1$. A network with a single node can be regarded as a node within its 0-step neighborhood, though our definition of LAP will not cover this case. Here we take $r$ as a positive integer which governs the type of LAP.

**2.2 Local AP-targeted attack (LAPTA)**

A common measure of the structural integrity of a network is the size of its giant connected component (GCC). APs are natural targets whose removal would disconnect the GCC if we aim for damaging the network immediately, and this method is called AP-targeted attack (APTA). In this paper, we propose a LAP-targeted attack method (LAPTA) which is like APTA (see Ref [4] for much more detail). The iterative process of LAPTA is as follows:

**Step-1:** For each node, find its maximum $r$-step ($r_m$) neighborhood where the current node can be regards as a LAP.

**Step-2:** For every $LAP^{r_m}$, calculate its 'destructivity', i.e. how many nodes will be disconnected within its $r_m$-step neighborhood after its removal.

**Step-3:** Rank all $LAP^{r_m}$s by their destructivity, and remove the most destructive one with all the links attaching to it. If the most destructive $LAP^{r_m}$ is not unique, then randomly choose one of them to remove.

**Step-4:** Repeat step 1 to 3 until the network does not contain any LAP.

The performance of the LAPTA strategy compared with the APTA strategy in a sample network was shown in Fig. 2.

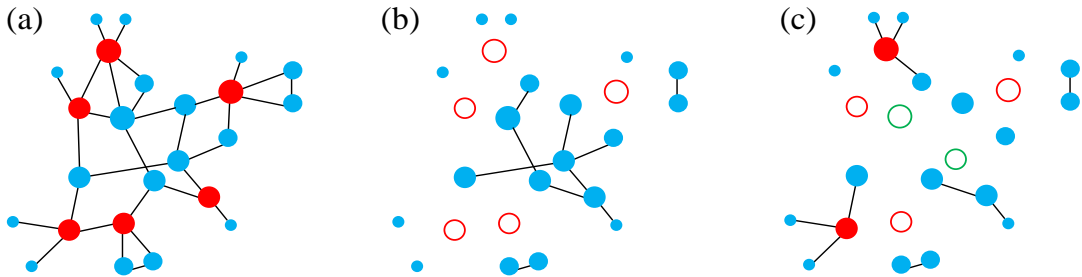

Fig. 2. Network demolition. (a) a sample network with twenty-four nodes and five varied degrees. The performance of network demolition when iteratively removing five nodes (green circle nodes and red circle nodes) under AP-targeted attack strategy and LAP-targeted attack strategy are respectively shown in (b) and (c).

When iteratively removing five nodes (red circles) in the network (a) with the strategy of APTA, we find the result as Fig. 2 (b) shows, in which just nine nodes stay unaffected. By contrast, when applying the LAPTA strategy to remove the five most influential nodes (three red circles and two green circles, as shown in Fig. 2 (c)), only four nodes are left in the largest connected component. In addition, it is worth to mention that the two green circle nodes in Fig. 2 (c) are not APs, corresponding to the same nodes in Fig. 2 (a), their destructivity has been counted by $LAP^1$ in the targeted attack process. The example in Fig. 2 illustrates that the LAPTA strategy is prior over APTA in terms of the effectiveness of network collapse.

# 3 Results

To examine the property of LAP in complex networks, we apply it to forty-five networks with eight various types as listed in Table 1-8 (see Appendix A). The networks are five infrastructure networks, ten autonomous systems networks, six peer-to-peer file sharing networks, five food webs networks, one neural network, one protein network, four communication networks, eight social networks, and five collaboration networks.

### 3.1 LAPs in real networks

The definition of LAP in the previous section prompts us to study the fraction of $LAP^r s$, $n_{LAP}^r$ ($n_{LAP}^r = N_{LAP}^r/N$), in real-world networks. Here $N_{LAP}^r$ and $N$ represent the number of $LAPs$ measured within each node's $r$-step neighborhood and the total number of nodes in a network, respectively. For easier comparison, we set $r$ under the same scale by dividing the diameter ($D$) of each network.

Firstly, let us focus on $n_{LAP^r}^{real}$ when $s^{real} < 0.1$ ($s = r/D$) as shown in Fig. 3 (a). Some real networks like infrastructure networks and file sharing networks, have a high proportion of LAPs. For example, the Beijing Road network (see the detail information in Appendix A) has almost the largest fraction (> 90%) of LAPs among all the real networks analyzed in this work, while social networks and collaboration networks which are rich of loop structures have a very small percentage (< 40%). Then with the increase of $s^{real}$, $n_{LAP}^{real}$ decreases. This may be caused by the appearance of complex connected structure, such as loop structure, thus some $LAP^r$s disappear with the increase of $r$.

Interestingly, we find that $n_{LAP}^{real}$ remains unchanged at a certain $s_{cr}^{real}$ for all the real networks investigated in this paper. The maximum $s_{cr}^{real}$ is equal to 0.5 as shown in figure 3 (b). The relationship between APs and LAPs tells us that $n_{LAP^{\lceil D*s_{cr}^{real}\rceil}}^{real}$ is the same as $n_{AP}^{real}$ when $s^{real} \geq s_{cr}^{real}$, which indicates that APs are the same as $LAP^{\lceil D*s_{cr}^{real}\rceil}$s. This result may be instructive to design a new algorithm for searching APs based on a depth-first search or a removal calculation. However, there is a special case for all the food webs which have no APs (see Table 4 in Appendix A). Meanwhile, these networks tend to be biconnected, and the extinction of one species would not disconnect the whole community [20]. Thus, aiming at removing APs to destroy food webs is not wise. However, the fraction of $LAP^1$s (average ~17.3%) in food webs networks we found is from average to 17.3%, and this may provide a new thought of network dismantling in this kind of network. In case a network has no APs, then the $s_{cr}^{real}$ of the network does not exist.

This regularity of $s_{cr}^{real}$ encourages us to explore the topological characteristics which may influence the quantity of $s_{cr}^{real}$. A comparison of $s_{cr}^{real}$ between a given real network with that of its randomized counterparts was made below. To achieve this, we firstly randomized each real network by a complete randomization procedure which turns the network into an Erdős − Rényi (ER) type of random network, with the number of nodes $N$ and links $L$ unchanged. The results showed that most completely randomized networks have a varied number of $s_{cr}^{rand-ER}$, corresponding to their initial networks.

For example, the quantity of $s_{cr}^{rand-ER}$s in some randomized networks whose initial networks are social networks or collaboration networks even become 0. This indicates that complete randomization process eliminates the topological characteristics which determine $s_{cr}^{real}$. By contract, when a degree-preserving randomization process is applied in the networks, rewiring the links among nodes to keep the degree $k$ of each node unchanged, then the $s_{cr}^{real}$ and $n_{AP}^{real}$[4] would not alter significantly as shown in figure 3 (d). There is no $s_{cr}^{rand-degree} = 0$, but $s_{cr}^{rand-degree}$ of a food web network is consistent with the results of its initial network. In other words, the characteristics of a network in terms of $s_{cr}^{real}$ is largely encoded in its degree distribution $p(k)$. However, most of the real-networks display smaller $s_{cr}^{real}$ than their degree-preserving randomized counterparts. We attribute these differences to higher-order structure correlations like clustering which may be eliminated in the degree-preserving randomization process. The maximum $s_{cr}^{rand-ER}$ and $s_{cr}^{rand-degree}$ are less than 0.5 or also equal to 0.5. It is similar to the results investigated in real-world networks in Fig. 3 (b).

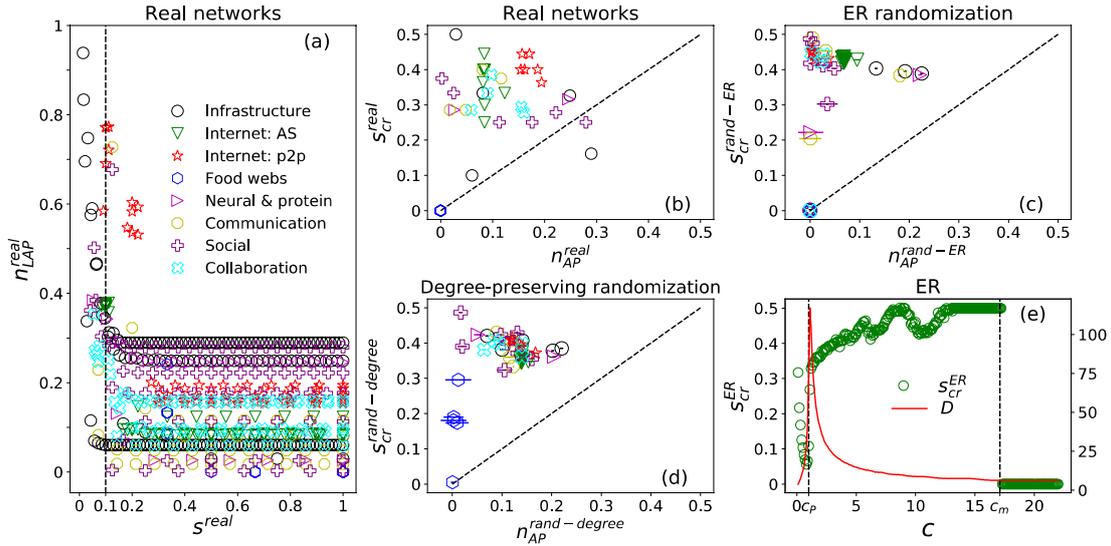

Fig. 3. LAPs in real networks and ER networks. (a) The fraction of local articulation points ($n_{LAP}^{real}$) versus radius percentage $s$ ($s = r/D$) is plotted for a wide range of real networks. (b) Critical radius fraction ($s_{cr}^{real}$) and relative size of $n_{LAP}^{real}$ obtained from the real networks. (c) Critical radius fraction ($s_{cr}^{rand-ER}$) and relative size of the fraction of local articulation points ($n_{LAP}^{rand-ER}$) obtained from the fully randomized counterparts of the real networks. (d) Critical radius fraction ($s_{cr}^{rand-degree}$) and relative size of the fraction of local articulation points ($n_{LAP}^{rand-degree}$) obtained from the fully randomized counterparts of the real networks. (e) $s_{cr}^{ER}$ versus different $c$ is plotted, where $c$ is the mean degree. The red line represents $D$ in different $c$. Simulations for $n_{LAP}^{r}$ are performed with network size $N = 10^5$, and the results are averaged over 100 realizations. In (c) and (d), all data points and error bars (standard error of the mean or s.e.m) are determined from 100 realizations of randomized networks. The dashed lines ($y = x$) are guided for eyes in (b-d). For detailed distribution of these real networks and their references, see Appendix A.

The results of $s_{cr}^{rand-ER}$ and $s_{cr}^{rand-degree}$ in Fig.3 (c) and (d) prompt us to

calculate $s_{cr}$ for networks with prescribed degree distribution. To achieve that, we count $s_{cr}^{ER}$ in ER random networks with Poisson degree distributions $p(k) = e^{-c}c^k/k!$, where $c$ is the mean degree. As shown in Fig. 3 (e), $s_{cr}^{ER}$ decreases with the increase of $c$ from 0 to $c_p$ and begins to increase after $c_p(c_P = 1)$, where $c_p$ is the critical point of ordinary percolation that emerges the GCC [21, 22].

For ER networks, the phenomenon can be explained as follows. The process of increasing the mean degree $c$ of an ER network can be considered as the process of randomly adding links into the network. When the mean degree $c$ is very small (nearly zero), there are only isolated nodes and dimers (components consisting of two nodes which connected by one link) and $n_{AP} \to 0$, which indicates that there is no $s_{cr}^{ER}$. Then, when $c$ gradually increase but still smaller than $c_P$, the network is full of finite connected components (FCCs), and most of them are trees. Hence, in the range of $0 < c < c_P$, most nodes (except isolated nodes and leaf nodes) are APs, and adding more links to the network will increase the number of APs. In another word, there is no difference between APs and LAPs. The value of $D$, which is shown in Fig. 3 (e) with the red solid line, also presents an increase, while the number of $s_{cr}^{ER}$ decreases. When $c \geq c_P$, the GCC develops and occupies a finite fraction of nodes in the network. In this case, there are two types of links to be added into the network: (I) links inside the GCC; and (II) links that connect the GCC with a FCC or two FCCs with the increase of $c$ [4, 5]. The probability of adding link-I is an increasing function, while type-II is a decreasing function [22]. Meanwhile, adding link-I to the network will never induce new APs, which means the results may convert the existing APs back to LAPs and reduce the diameter. Conversely, adding link-II will never decrease the number of APs, and normal nodes can also be converted into APs and may increase the diameter. The contributions of these two types of links to $D$ compete, and link-I plays a more important role. These processes make $s_{cr}^{ER}$ start to increase. Interestingly, the maximum $s_{cr}^{ER}$ is equal to 0.5 with the increase of $c$, although the values of $s_{cr}^{ER}$ fluctuate due to the complex competition of two types of links. When APs disappear, there is no $s_{cr}^{ER}$ after $c_m$ ($c_m \approx 17.1$).

Our results illustrate that the APs of a network can be identified by the local search, not global search. LAP$^{[D*s_{cr}]}$s can be an efficient alternative to APs, for not having to know the global information of a network when establish an adjacency matrix of a network. Meanwhile, this discovery can be a novel potential approach to find APs quickly in large-scale dynamical networks, such as social networks though parallel computing technology.

**3.2 Network dismantling**

An enduring truth of network science is that the removal of a few highly connected nodes, or hubs, can break up a complex network into many disconnected components [23]. The dismantling of complex networks is underlying many important applications in network science, in terms of optimal, vaccination and surveillance, information spreading, viral marketing, and identification of influential nodes[24]. Considerable research efforts have been devoted to the network dismantling problems recently [25-28]. Representing natural vulnerabilities of a network, APs are potential targets of attack if one aims for immediate damage to a network. A brute-force AP targeted attack (APTA) was proposed to make most nodes disconnected from the GCC of the current network. While the algorithm's effectiveness is limited by "a small fraction of nodes to be removed" [4]. In fig. 4 we compute the relative size of the GCC as $N_{gcc}$, which is a function refers the fraction of removed nodes $p$ by using LAPTA and APTA destruction strategies in real networks (Fig. 4-A to H) and ER networks (Fig. 4-I).

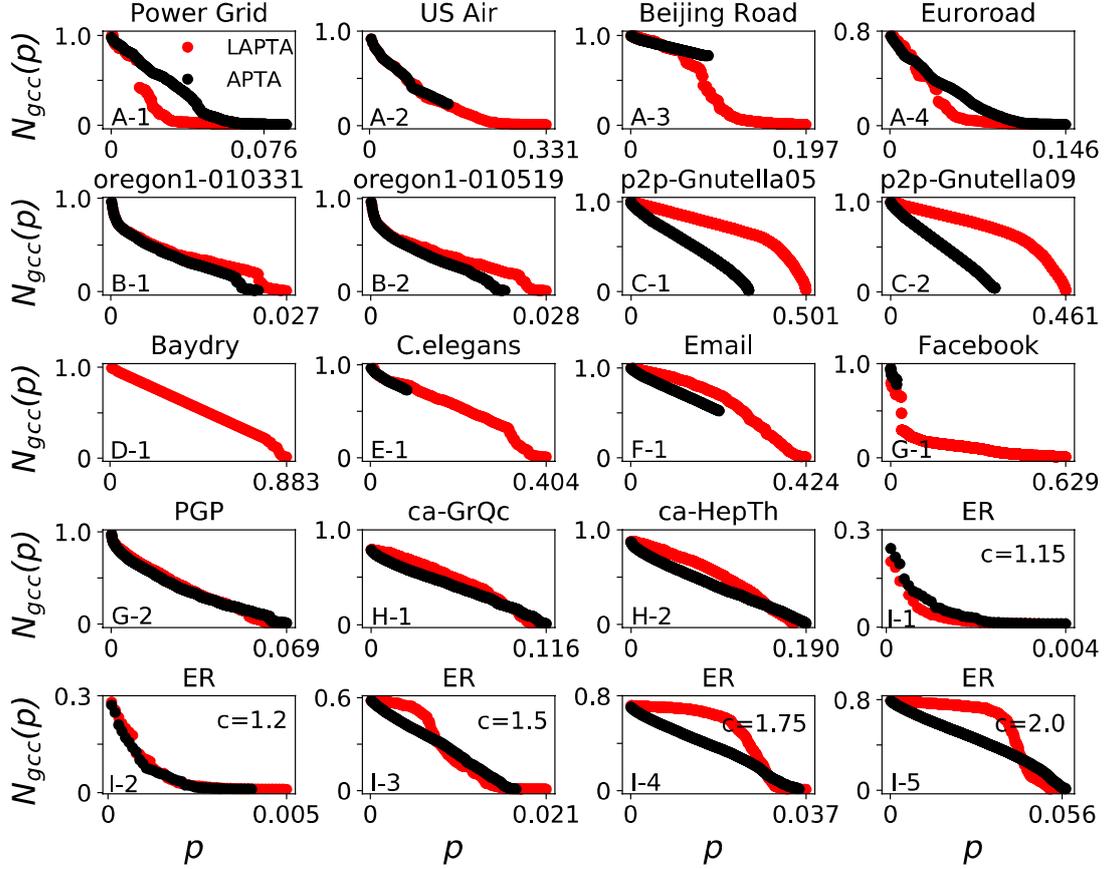

Fig. 4. Comparison of different network destruction strategies by reducing the size of GCC for both real networks (A-H) and model networks (I). The real networks include Infrastructure networks (Power Grid (A-1), US Air (A-2), Beijing Road (A-3), Euroroad (A-4)), autonomous systems (oregon1-010331(B-1), oregon1-010519 (B-2)), peer-to-peer file sharing networks (p2p-Gnutella05 (C-1), p2p-Gnutella09 (C-2)), food webs (Baydry (D-1)), neural & protein network (C.elegans (E-1)), communication network (Email (F-1)), social network (Facebook (G-1), PGP (G-2)) and collaboration networks (ca-GrQc (H-1), ca-HepTh (H-2)). The data of model networks are determined from 100 independent network instances of size $N = 10^5$.

The experiments show that, for infrastructure networks, removing the set of influencers identified by the LAPTA algorithm is more effective than removing the APs that are identified through APTA owing to their spatially embedded structures. The infrastructure networks are often organized by embedding nodes and edges in space. Due to the high cost of adding new links such as power grids or roads to their corresponding networks, a node always plays an indispensable role for the local connection. So $c$ is always smaller than other type of networks like communication networks. AP removal strategy only maximizes the damage to the network in every step which starts from the periphery of the network. While LAP removal strategy may attack the core of a network, even though it may not reduce GCC immediately. After a certain level of network destruction, GCC may show a sharp drop by LAPTA strategy as shown in Fig. 4 (A-1, A-3 and A-4). This phenomenon can also be found in modeling networks as shown in Fig.4 (I-1 to I-5). LAPTA strategy would gradually loses its efficiency with the increase of $c$, which suggested that the efficiency of LAPTA strategy relates $c$ closely. But US Air who possesses a very high $c$ is an exception, we own this phenomenon to assortativity and clustering characteristics of airline networks [29]. Moreover, in social networks, such as the Facebook networks and PGP networks, LAPTA strategy is also optimal. Networks, like US air, Baydry, C.elegans, Email and so on(see Appendix A), which have a large RGB can be destroyed completely using the LAPTA strategy, while the APTA strategy is useless in these networks after a limited fraction of nodes removed. However, APTA is still very efficient for autonomous systems, peer-to-peer file sharing networks and collaboration networks (see Fig. 4 B, C, and H).

## 4 Discussion

In this article, we extend the concept of AP to define a new type of node called local AP (LAP) by different $r$-step neighborhood. For discussing the relationship between LAPs and APs, we investigate the fraction of LAPs, $n_{LAP}^{real}$, in forty-five real networks, and a maximum percentage of $r$, $s_{cr}^{real}$, has been found. This regularity then could also be turned out in ER networks and degree-preserving networks produced by randomizing real-networks. Moreover, we also find that with the increase of $c$, $s_{cr}^{ER}$ decrease and has an obvious transition after $c = 1$, and the maximum $s_{cr}^{ER}$ in all range of c is also equal to 0.5. These simulation results have verified the possibility for developing a new search algorithm method of APs.

APs are important in ensuring the robustness and connectivity of many real-world networks. Representing natural vulnerabilities of a network, APs are potential targets of attack if one aims for immediate damage to a network. Given a limited 'budget' (the number of nodes to be removed), the APTA strategy is effective in reducing the GCC [4]. However, we find that LAPs are more common in both real-networks and model networks, even in some special networks like food web networks which nearly have no APs but possess 17.3% of LAP[1] on average.

This prompts us to design a LAPTA strategy by iteratively removing the most destructive LAP that whose removal could cause the most nodes disconnected from the local environment. It is worth to mention that the LAP targeted attack may not instantly reduce the GCC apparently. Indeed, at the beginning of dismantling network, we find that the performance of LAPTA is worse than the performance of APTA in reducing GCC in a wide range of real-networks as shown in Fig. 4. Nevertheless, after a small fraction of removed nodes, LAPTA becomes more efficient in reducing the GCC, especially in some infrastructure networks shown in Fig. 4 (A). Further, a residual GCC (RGB) is resulted by APTA in a network which depends on the structure of the current network. LAPTA would completely decompose the network as shown in Fig. 4 (A-2, E-1, F-1, G-1).

Despite this paper find a maximum percentage of $r$ for better searching the APs and design a more efficiently targeted attack strategy by LAPs, it lacks a theoretical explanation of the empirical findings. In the future work, we will try to investigate the fraction of $LAP^r$ in model networks and provide a theoretical explanation of the empirical findings about why the maximum percentage of $r$ is 0.5 in both the real-networks and the model networks. In addition, we will develop a more efficient method for dismantling network by combining the LAPTA and the APTA.

# Acknowledgments

The authors thank for support from the National Natural Science Foundation of China (No.71571017, No.91646124, and No.71621001), and support from the Fundamental Research Funds for the Central Universities (2015JBM058).

# Appendix A

All the real-world networks analyzed in this work are listed in tables (see Appendix A). The type, name and reference, number of nodes ($N$) and edges ($E$), and a brief description of all networks are included.

Table 1. Infrastructure network

| Network name | $N$ | $E$ | $p(RGB)$ | Description |
|---|---|---|---|---|
| US air lines [30] | 332 | 2,126 | 0.23193 | US airport network in 1997 |
| Chinese air lines [31] | 203 | 1819 | 0.00985 | Chinese airport network in 2015 |
| Power grid [32] | 4,941 | 6,594 | 0.00425 | Power grid in the Western Stated in US |
| Euroroad [33] | 1174 | 1417 | 0.00511 | A road network mostly in Europe |
| Beijing road | 3129 | 5539 | 0.80729 | Beijing road network in 2008 |

Table 2. Internet: Autonomous Systems (AS) [34]

| Network name | $N$ | $E$ | $p(RGB)$ | Description |
|---|---|---|---|---|
| as19971108 | 3,015 | 5,539 | 0.00066 | AS graph from November 08 1997 |
| oregon1-010331 | 10,670 | 22,002 | 0.00028 | AS peering information inferred from Oregon route-views |
| oregon1-010407 | 10,729 | 22,000 | 0.00037 | Same as above (at different time) |
| oregon1-010414 | 10,790 | 22,470 | 0.00037 | Same as above (at different time) |
| oregon1-010421 | 10,859 | 22,748 | 0.00027 | Same as above (at different time) |
| oregon1-010428 | 10,886 | 22,494 | 0.00027 | Same as above (at different time) |
| oregon1-010505 | 10,943 | 22,608 | 0.00027 | Same as above (at different time) |
| oregon1-010512 | 11,011 | 22,678 | 0.00027 | Same as above (at different time) |
| oregon1-010519 | 11,051 | 22,725 | 0.00027 | Same as above (at different time) |
| oregon1-010526 | 11,174 | 23,410 | 0.00026 | Same as above (at different time) |

Table 3. Internet: peer-to-peer (p2p) file sharing networks [35, 36]

| Network name | $N$ | $E$ | $p(RGB)$ | Description |
|---|---|---|---|---|
| p2p-Gnutella04 | 10,876 | 39,994 | 0.00055 | Gnutella p2p file sharing network |
| p2p-Gnutella05 | 8,846 | 31,839 | 0.00045 | Same as above (at different time) |
| p2p-Gnutella06 | 8,717 | 31,525 | 0.00002 | Same as above (at different time) |
| p2p-Gnutella08 | 6,301 | 20,777 | 0.00032 | Same as above (at different time) |
| p2p-Gnutella09 | 8,114 | 26,013 | 0.04363 | Same as above (at different time) |
| p2p-Gnutella24 | 26,518 | 65,369 | 0.000189 | Same as above (at different time) |

Table 4. Food webs [18]

| Network name | $N$ | $E$ | $p(RGB)$ | Description |
|---|---|---|---|---|
| Baydry | 128 | 2,137 | 1 | Food webs at Florida Bay, Dry Season |
| Baywet | 128 | 2,106 | 1 | Food webs at Florida Bay, wet Season |
| Everglades | 69 | 916 | 1 | Food webs at Everglades Graminoid Marshes; Wet Season |
| Narragan | 35 | 220 | 1 | Food webs at Narragansett Bay |
| StMarks | 54 | 356 | 1 | Food webs at Marks River |

Table 5. Neural & protein network

| Network name | $N$ | $E$ | $p(RGB)$ | Description |
|---|---|---|---|---|
| C.elegans [37] | 453 | 2,025 | 0.72848 | Neural network of C. elegans |
| Protein [38] | 1870 | 2277 | 0.003743 | Protein interaction network |

Table 6. Communication network

| Network name | $N$ | $E$ | $p(RGB)$ | Description |
|---|---|---|---|---|
| Enron [18] | 69,244 | 255,984 | 0.00687 | Enron email communications |

| Email-EuAll [35] | 265,214 | 420,045 | 0.000512 | Email network from a large European research institute |
| Email [39] | 1,133 | 5,451 | 0.52251 | E-mail network of URV |
| Terrorist [40] | 62 | 152 | 0.46774 | The terrorist communication network of the 9/11 attacks on U.S |

Table 7. Social network

| Network name | $N$ | $E$ | $p(RGB)$ | Description |
|---|---|---|---|---|
| Epinions [18] | 131,828 | 711,783 | 0.00232 | Who-trusts-whom network of Epinions.com |
| Facebook [41] | 4039 | 88,234 | 0.77792 | Social circles from Facebook |
| loc-Gowalla [42] | 196,591 | 950,327 | 0.00846 | Gowalla location based online social network |
| loc-Brightkite [42] | 58,228 | 214,078 | 0.00194 | Brightkite location based online social network |
| Jazz [43] | 198 | 2,742 | 0.94949 | Jazz musicians netwrok |
| Karate club [44] | 34 | 78 | 0.14706 | Zachary's Karate Club |
| Dolphins [45] | 62 | 159 | 0.64516 | Lusseau's Bottlenose Dolphins |
| PGP [46] | 10,680 | 24,316 | 0.00562 | Shares confidential information using the Pretty Good Privacy |

Table 8. Collaboration network [35]

| Network name | $N$ | $E$ | $p(RGB)$ | Description |
|---|---|---|---|---|
| ca-AstroPh | 18,772 | 198,110 | 0.65667 | Collaboration network of Arxiv Astro Physics |
| ca-Condmat | 23,113 | 93,497 | 0.25608 | Collaboration network of Arxiv Condensed Matter |
| ca-GrQc | 5,242 | 14,496 | 0.00744 | Collaboration network of Arxiv General Relativity |
| ca-HepPh | 12,008 | 118,521 | 0.21677 | Collaboration network of Arxiv High Energy Physics |
| ca-HepTh | 9,877 | 25,998 | 0.00294 | Collaboration network of Arxiv High Energy Physics Theory |